\documentclass[onecolumn,showpacs]{revtex4}
\usepackage{graphicx}
\usepackage{dcolumn}
\usepackage{amsmath}
\usepackage{amssymb}
\usepackage{hyperref}
\usepackage{scalerel}

\allowdisplaybreaks
\newtheorem{theorem}{Theorem}[section]

\newtheorem{proposition}[theorem]{Proposition}

\newcommand{\qed}{\nobreak \ifvmode \relax \else
	\ifdim\lastskip<1.5em \hskip- \lastskip
	\hskip1.5em plus0em minus0.5em \fi \nobreak
	\vrule height0.75em width0.5em depth0.25em\fi}

\begin{document}

\title{Homothetic Killing vectors in stationary axisymmetric vacuum spacetimes}

\author{Abbas \surname{Sherif}}
\email{abbasmsherif25@gmail.com}
\affiliation{Cosmology and Gravity Group, Department of Mathematics and Applied Mathematics, University of Cape Town, Rondebosch 7701, South Africa}

\author{Peter K. S. \surname{Dunsby}}
\email{peter.dunsby@uct.ac.za}
\affiliation{Cosmology and Gravity Group, Department of Mathematics and Applied Mathematics, University of Cape Town, Rondebosch 7701, South Africa\\South African Astronomical Observatory, Observatory 7925, Cape Town, South Africa}

\author{Rituparno \surname{Goswami}}
\email{vitasta9@gmail.com}
\affiliation{Astrophysics and Cosmology Research Unit, School of Mathematics, Statistics and Computer Science, University of KwaZulu-Natal, Private Bag X54001, Durban 4000, South Africa}

\author{Sunil D.\ \surname{Maharaj}}
\email{Maharaj@ukzn.ac.za}
\affiliation{Astrophysics and Cosmology Research Unit, School of Mathematics, Statistics and Computer Science, University of KwaZulu-Natal, Private Bag X54001, Durban 4000, South Africa}

\begin{abstract}

In this paper we consider homothetic Killing vectors in the class of stationary axisymmetric vacuum (SAV) spacetimes, where the components of the vectors are functions of the time and radial coordinates. In this case the component of any homothetic Killing vector along the \(z\) direction must be constant. Firstly, it is shown that either the component along the radial direction is constant or we have the proportionality \(g_{\phi\phi}\propto g_{\rho\rho}\), where \(g_{\phi\phi}>0\). In both cases, complete analyses are carried out and the general forms of the homothetic Killing vectors are determined. The associated conformal factors are also obtained. The case of vanishing twist in the metric, i.e., \(\omega= 0\) is considered and the complete forms of the homothetic Killing vectors are determined, as well as the associated conformal factors. 
\end{abstract}


\maketitle

\section{Introduction}	

The existence of conformal symmetries in spacetimes has extensive applications in generating new solutions to the Einstein field equations, as well as simplifying solutions to the field equations (i.e., metrics solving the field equations) \cite{ac1,ac2,ac3,ac4,her1,her2,pet1,mt1}. Conformal symmetries, implied by the existence of conformal Killing vector fields (vector fields which preserve the metric up to scale when the metric is Lie dragged along the vector field), also provides information on the kinematical quantities of the spacetime by specifying restrictions on these quantities \cite{rm3,rm4,rm1,rm2,ac1}. Special cases of conformal Killing vector fields - homothety (where the conformal factor is constant), special homothety (where the covariant derivative of the conformal factor is constant), and Killing vector field (where the conformal factor vanishes), all have interesting properties that have proven useful in the study of geodesic motions in spacetimes. The case of homothety, whose existence implies the self-similarity of a spacetime has also been studied extensively by various author, and consequently its implications during gravitational collapse have been explored (for example see the references \cite{c1,c2,br1,rob1,chr1,chr2,chr3,chr4}). In some applications the geometry of the spacetime, in particular the sign of the Ricci curvature tensor can specify the nature of the symmetry induced by the Killing vector fields. The case of conformal symmetries in locally rotationally symmetric spacetimes was studied by Apostolopoulos and Tsamparlis \cite{mt2}, and Tsamparlis \textit{et al.} \cite{mt3} found the proper conformal Killing vector fields, homothetic Killing vector fields and special homothetic Killing vector fields for Bianchi I class of spacetimes.

The line element \eqref{sav10} of stationary axisymmetric vacuum (SAV) spacetimes represents a general vacuum spacetime around a central body with arbitrary multipole moments. This solution to the Einstein field equations arises naturally from the solution to the Ernst equation with potential, denoted \(\mathcal{E}\), satisfying \eqref{sav11} \cite{ep1}. Symmetries of \eqref{sav10} provide a means of generating new solutions which come with a set of integrability conditions. In general, the equations of motion arising from \eqref{sav10} are not integrable. This is seen for a special subclass of \eqref{sav10} known as the Zipoy-Voorhees (ZV) metric \cite{zi1,vo1} parametrized by a real constant \(\delta\). The SAV metric \eqref{sav10} also generalizes some well known solutions such as the Manko-Novikov \cite{mn1} and the QM \cite{hq1,hq2,orl1} solutions, which in turn are generalizations of the well known Kerr solution for specified restriction on the quadropole parameter.

Various authors have employed varying approaches to investigate integrability of the equations of motion in the ZV spacetime \cite{mv1,jb1,lg1,ni1}. From the general Killing equation (including up to higher orders) \(T^{(\alpha\dots\beta;\gamma)}=0\), it can be shown that the full contraction of an arbitrary symmetric \(n\)-tensor by the momentum of a particle in motion \(T^{(\alpha\dots\beta)}p_{\alpha}\dots p_{\beta}\) is a constant (called a first integral) of along the geodesic of the particle, and thus gives a constant of motion (see the reference \cite{jb2} for more discussion). The independence of SAV spacetimes on the \(t\) and \(\phi\) coordinates provides two constants of motion from the two Killing vectors \(\partial_t\) and \(\partial_{\phi}\) respectively. A third constant of motion is the rest mass, obtained by choosing the symmetric tensor of the Killing equation as the metric \(g^{\alpha\beta}\). Obtaining the fourth constant of motion would indicate complete integrability, and allow for solution by quadrature. This task was rigorously explored, both analytically and numerically, in \cite{mv1,jb2,ni1,jb3}. Kugrlikov and Metveev \cite{mv1} excluded the possibility of finding a first integral as polynomial in momenta up to degree \(6\) for the case \(\delta=2\). The same case was considered by Maciejewski \textit{et al.} \cite{ni1}, where the authors showed that the identity component of the differential Galois group of the variational and new variational equations associated to the equations of motion was non-abelian, which implies non-integrability \cite{am1,mr1}. This proof ruled the existence of a much wider class of functions than that considered by \cite{mv1}, i.e., the class of \textit{meromorphic} functions . A detailed numerical study done by Lukes-Gerakolopoulos \cite{lg1} showed that away from the Schwarzschild's limit \(\delta=1\), geodesic motion was unstable and there was a breakdown in the predictability of the orbits and thus further confirmed the results of \cite{ni1}. Away from the Schwarzschild's case, i.e., \(1<\delta\leq2\), this region of instability was quantified via the computation of the Arnold tongue of instability in \cite{as1}. Brink, through efforts to find first integrals, developed sophisticated analytic algorithms to find second and fourth order Killing tensors (the latter being the first attempt at that order) \cite{jb2,jb3}. 

A more recent physical application involving the Zipoy-Voorhees metric was carried out in \cite{orl2}, where the authors studied neutrino oscillations in the field of the spacetime by determining the phase shift generating small deformations.

In this paper we study homothetic Killing vectors in SAV spacetimes restricting the components of the vectors to be dependent on \(t\) and \(\rho\) coordinates only. The paper has the following structure: Section \ref{soc1} provides the definition and brief discussion of the SAV  spacetimes. In Section \ref{soc12} we investigate Homothetic Killing vectors in SAV spacetimes, under certain restriction on their component functions. We also consider the particular case where there is no twisting term in the metric. Finally, we conclude with discussion of our results in Section \ref{soc6}.

\section{SAV spacetimes}\label{soc1}

We begin this section by introducing the class of SAV spacetime and computing quantities from the metric components, to be used in subsequent calculations. 

The line element of a general SAV spacetime is given by 

\begin{eqnarray}\label{sav10}
\begin{split}
\underline{g}&=k^2e^{-2\psi}\left[e^{2\gamma}\left(d\rho^2+dz^2\right)+R^2d\phi^2\right]-e^{2\psi}\left(dt-\omega d\phi\right)^2,
\end{split}
\end{eqnarray}
with the metric functions

\begin{eqnarray*}
\begin{split}
\underline{g}_{tt}=-\frac{1}{\omega}\underline{g}_{\phi t}=-e^{2\psi};\ \ \underline{g}_{\phi\phi}=W;\ \ \underline{g}_{\rho\rho}=\underline{g}_{zz}=k^2e^{2\left(\gamma-\psi\right)},
\end{split}
\end{eqnarray*}
where we have defined 

\begin{eqnarray}\label{micp}
W=k^2R^2e^{-2\psi}-\omega^2e^{2\psi}. 
\end{eqnarray}
Here \(k\in \mathbb{R}, \psi=\psi\left(\rho,z\right), \gamma=\gamma\left(\rho,z\right),\omega=\omega\left(\rho,z\right)\) and \(R=R\left(\rho,z\right)\) (see references \cite{ep1,jb1} for details). These spacetimes are constructed from a complex Ernst potential \(\mathcal{E}\) \cite{ep1} via the equation

\begin{eqnarray}\label{sav11}
\mathfrak{R}\left(\mathcal{E}\right)\bar{\nabla}^2\mathcal{E}=\bar{\nabla}\mathcal{E}\cdot\bar{\nabla}\mathcal{E},
\end{eqnarray}
where \(\mathfrak{R}\left(\mathcal{E}\right)=e^{2\psi}\) is the real part of \(\mathcal{E}\), \(\bar{\nabla}^2=\partial_{\rho\rho}+\left(1/\rho\right)\partial_{\rho}+\partial{zz}\), \(\bar{\nabla}=\left(\partial_{\rho},\partial_z\right)\). The functions \(\gamma\) and \(\omega\) can be obtained from line integrals of \(\mathcal{E}\), and the function \(R\) solves

\begin{eqnarray}\label{rhoeq}
R_{,\rho\rho}+R_{,zz}=0,
\end{eqnarray}
where the ``comma" denotes partial differentiation. The independent non-zero Christoffel symbols are given by 

\begin{subequations}
\begin{align}
\Gamma^k_{jA}&=\frac{1}{2}\underline{g}^{ik}\underline{g}_{ij,A};\ \ \Gamma^{\rho}_{\rho\rho}=-\Gamma^{\rho}_{zz}=\Gamma^{z}_{z\rho}=\frac{1}{2}\underline{g}^{\rho\rho}\underline{g}_{\rho\rho,\rho},\ \ \left(i,j,k\in\lbrace{t,\phi\rbrace};\ \ A\in\lbrace{\rho,z\rbrace}\right)\;, \label{eq29}\\
\Gamma^A_{jk}&=-\frac{1}{2}\underline{g}^{\rho\rho}\underline{g}_{jk,A};\ \ \Gamma^{z}_{zz}=-\Gamma^{z}_{\rho\rho}=\Gamma^{\rho}_{z\rho}=\frac{1}{2}\underline{g}^{\rho\rho}\underline{g}_{\rho\rho,z}\ \ \left(j,k\in\lbrace{t,\phi\rbrace};\ \ A\in\lbrace{\rho,z\rbrace}\right).\label{eq30}
\end{align}
\end{subequations}

\section{The conformal Killing equations and homothetic Killing vectors in SAV spacetimes}\label{soc12}

For an arbitrary vector field \(v\), the Lie derivative of the metric along \(v\) is given by

\begin{eqnarray}\label{sav30}
\mathcal{L}_v\underline{g}_{\mu\nu}=2\Psi \underline{g}_{\mu\nu}.
\end{eqnarray}
The function \(\Psi\) determines if \(v\) is a Killing vector (KV), a homothetic Killing vector (HKV), a special homothetic Killing vector (SHKV) or \textit{proper} conformal Killing vector (CKV) field:

\begin{eqnarray*}
\begin{split}
\Psi&=0\implies v\ \ KV,\\
\Psi&=const.\neq 0 \implies v\ \ HKV,\\
\Psi&=nonconst.\implies v\ \ \text{\textit{proper}}\ \ CKV.
\end{split}
\end{eqnarray*}
In terms of the covariant derivative \eqref{sav30} may be written as

\begin{eqnarray}\label{sav31}
v_{(\mu;\nu)}=\Psi \underline{g}_{\mu\nu},
\end{eqnarray}
where we have used the ``semicolon" to denote the covariant derivative, and the parenthesis is the usual symmetrization of the indices.

We shall seek homothetic Killing vectors of the form

\begin{eqnarray}\label{sav32}
\eta=C_1\partial_t+C_2\partial_{\rho}+C_3\partial_{z}+C_4\partial_{\phi}.
\end{eqnarray}
We shall consider cases for which we have the components \(C_i\) for \(i\in\lbrace{1,2,3,4\rbrace}\) are functions of \(t\) and \(\rho\) only. Using 

\begin{eqnarray}\label{eq35}
\nabla_{(\mu}\eta_{\nu)}=\Psi \underline{g}_{\mu\nu},
\end{eqnarray}
where \(\Psi\) is the conformal factor, the set of CKEs become

\begingroup
\begin{subequations}
\begin{align}
0&=C_{1,t}-\Psi-\omega C_{4,t}+\psi_{,\rho}C_2+\psi_{,z}C_3,\label{eq101}\\
0&=\left(C_{1,\rho}-\omega C_{4,\rho}\right)e^{2\psi}-k^2C_{2,t}e^{2\left(\gamma-\psi\right)},\label{eq102}\\
0&=C_{3,t},\label{eq103}\\
0&=WC_{4,t}+\biggl[C_2\left(\omega_{,\rho}+2\omega\psi_{,\rho}\right)+C_3\left(\omega_{,z}+2\omega\psi_{,z}\right)+\omega\left(C_{1,t}-2\Psi\right)\biggr]e^{2\psi},\label{eq104}\\
0&=C_{2,\rho}+C_2\left(\gamma-\psi\right)_{,\rho}+C_3\left(\gamma-\psi\right)_{,z}-\Psi,\label{eq105}\\
0&=C_{3,\rho},\label{eq106}\\
0&=WC_{4,\rho}+\omega C_{1,\rho}e^{2\psi},\label{eq107}\\
0&=C_3\left(\gamma-\psi\right)_{,z}+C_2\left(\gamma-\psi\right)_{,\rho}-\Psi,\label{eq108}\\
0&=W_{,\rho}C_2+W_{,z}C_3-2W\Psi.\label{eq109}
\end{align}
\end{subequations}
\endgroup

From \eqref{eq103} and \eqref{eq106} we have that \(C_3=l\) where \(l\) is an arbitrary constant. Combining \eqref{eq105} and \eqref{eq108} we see that \(C_2=C_2\left(t\right)\).

Taking the time derivative of the constraint equations \eqref{eq108} and \eqref{eq109}, and comparing we have

\begin{eqnarray}\label{eq120}
C_{2,t}\left(W_{,\rho}-2W\left(\gamma-\psi\right)_{,\rho}\right)=0.
\end{eqnarray}
So either \(C_{2,t}=0\) in which case \(C_2=m=constant\), or 

\begin{eqnarray}\label{eq121}
W_{,\rho}-2W\left(\gamma-\psi\right)_{,\rho}=0.
\end{eqnarray}
Hence, we have the following proposition.

\begin{proposition}\label{lem2}
Assume that the metric \eqref{sav10} admits HKVs of the form \eqref{sav32}, where \(C_i=C_i\left(t,\rho\right)\) for \(i\in\lbrace{1,2,3,4\rbrace}\). Then \(C_3\) must be constant, and either

\begin{enumerate}
\item \(C_2\) is constant, in which case \(C_4=C_4\left(t\right)\); or

\item \(C_2=C_2\left(t\right)\) and \(\underline{g}_{\phi\phi}=k^{-2}f\underline{g}_{\rho\rho}\),
\end{enumerate}
for an arbitrary positive function \(f=f\left(z\right)\).
\end{proposition}

We will now consider the 2 cases of Proposition \ref{lem2} in detail.

\subsection{The case of constant \(C_2\)}

Noting that \(C_2=m=constant\), we have the conformal factor \(\Psi\) as

\begin{eqnarray}\label{eqmic}
\Psi=l\left(\gamma-\psi\right)_{,z}+m\left(\gamma-\psi\right)_{,\rho},
\end{eqnarray}
in which case we see that \(\Psi=\Psi\left(\rho,z\right)\). Furthermore, \eqref{eq102} and \eqref{eq107} reduce respectively to 

\begin{subequations}
\begin{align}
0&=C_{1,\rho}-\omega C_{4,\rho},\label{eq122}\\
0&=WC_{4,\rho}+\omega C_{1,\rho}e^{2\psi},\label{eq123}
\end{align}
\end{subequations}
which can be combined to give

\begin{eqnarray}\label{eq124}
C_{4,\rho}\left(W+\omega^2e^{2\psi}\right)=0.
\end{eqnarray}
The determinant \(\det\left(\underline{g}\right)=-e^{2\psi}\underline{g}_{\rho\rho}^2\bar{W}\), where \(\bar{W}=W+\omega^2e^{2\psi}\). Therefore we rule out the case \(\bar{W}=0\), since otherwise \(\underline{g}\) is not invertible. Hence we must have \(C_4=C_4\left(t\right)\). 

Suppose we have that \(C_4=C_4\left(t\right)\). Then from \eqref{eq122} we have \(C_1=C_1\left(t\right)\). Substituting \eqref{eq101} into \eqref{eq104} and using \eqref{eqmic} we have the following differential equation for \(C_4\):

\begin{eqnarray}\label{eqq1}
C_{4,t}=-\frac{Me^{2\psi}}{W+\omega^2e^{2\psi}},
\end{eqnarray}
which gives

\begin{eqnarray}\label{eqq2}
C_4=-\left(\frac{Me^{2\psi}}{W+\omega^2e^{2\psi}}\right)t+a_1,
\end{eqnarray}
(note that the parenthesized term of \eqref{eqq2} is constant since all quantities therein are independent of \(t\)) for an arbitrary constant \(a_1\), where

\begin{eqnarray}\label{eqq3}
\begin{split}
M\left(\rho,z\right)&=m\left(\omega_{,\rho}+2\omega\psi_{,\rho}\right)+l\left(\omega_{,z}+2\omega\psi_{,z}\right)-\omega\left(l\gamma_{,z}+m\gamma_{,\rho}\right).
\end{split}
\end{eqnarray}
Furthermore, we combine \eqref{eqmic} and \eqref{eq101} to obtain

\begin{eqnarray}\label{eqq4}
\left(C_1-\omega C_4\right)_{,t}=\tilde{M},
\end{eqnarray}
which has the solution

\begin{eqnarray}\label{eqq5}
C_1-\omega C_4=\tilde{M}t+a_2,
\end{eqnarray}
for an arbitrary constant \(a_2\), where

\begin{eqnarray}\label{eqq6}
\tilde{M}\left(\rho,z\right)=l\left(\gamma-2\psi\right)_{,z}+m\left(\gamma-2\psi\right)_{,\rho}.
\end{eqnarray}
Hence we may write \(C_1\) explicitly as

\begin{eqnarray}\label{eqq7}
C_1=\left(\tilde{M}-\frac{\omega Me^{2\psi}}{W+\omega^2e^{2\psi}}\right)t+a_1\omega+a_2.
\end{eqnarray}
However, since \(C_1\) and \(C_4\) are functions of \(t\) and \(\rho\) only, we must have 

\begin{subequations}
\begin{align}
\frac{Me^{2\psi}}{W+\omega^2e^{2\psi}}&=\alpha^2,\label{eqq8}\\
\tilde{M}&=\beta,\label{eqq9}\\
\omega&=\pi,\label{eqq10}
\end{align}
\end{subequations}
for constants \(\alpha,\beta,\pi\), where \(\alpha,\beta,\pi\neq 0\) (the choice of the square on \(\alpha\) is just to keep the function \(e^{2\psi}\) without the square root). Using \eqref{micp} and simplifying \eqref{eqq8} we have that

\begin{eqnarray}\label{eqq11}
e^{4\psi}=k^2\alpha^2 \frac{R^2}{M}.
\end{eqnarray}

Now, suppose we have that \(l,m\neq 0\). As \(\omega=\pi=constant\), \eqref{eqq3} reduces to 

\begin{eqnarray}\label{eqq12}
M=-\pi\left[l\left(\gamma-2\psi\right)_{,z}+m\left(\gamma-2\psi\right)_{,\rho}\right].
\end{eqnarray}
We can always choose \(\beta=-1/\pi\) so that \(M=1\) (using \eqref{eqq9}), and hence from \eqref{eqq8} we have

\begin{eqnarray}\label{eqq13}
e^{2\psi}=k\alpha R.
\end{eqnarray}

Setting \(M=1\), \eqref{eqq12} has the general solution

\begin{eqnarray}\label{eqq14}
\gamma-2\psi=-\frac{1}{\pi m}\rho+r\left(z-\frac{l}{m}\rho\right),
\end{eqnarray}
for any smooth function \(r\left(z-\frac{l}{m}\rho\right)\), which gives

\begin{eqnarray}\label{eqq15}
e^{2\gamma}=k^2\alpha^2R^2e^{-2\left(\frac{1}{\pi m}\rho-r\left(z-\frac{l}{m}\rho\right)\right)}.
\end{eqnarray}

We therefore have the component \(C_i=C_i\left(t,\rho\right)\) as

\begin{subequations}
\begin{align}
C_1&=\pi_1t+\pi_2,\label{eqq16}\\
C_2&=m,\label{eqq17}\\
C_3&=l,\label{eqq18}\\
C_4&=-\alpha^2t+a_1,\label{eqq19}
\end{align}
\end{subequations}
where

\begin{eqnarray}\label{hellothere}
\begin{split}
\pi_1&=\frac{1}{\pi}-\pi\alpha^2,\\
\pi_2&=a_1\pi+a_2,
\end{split}
\end{eqnarray}
with metric functions \(\psi,\gamma\) and \(\omega\), given respectively in \eqref{eqq13}, \eqref{eqq15} and \eqref{eqq10}. The associated conformal factor can be written as
 
\begin{eqnarray}\label{eqq20}
\Psi_{m,l\neq 0}=\frac{1}{2k\alpha R}\left(mR_{,\rho}+lR_{,z}\right)-\frac{1}{\pi}.
\end{eqnarray}

To treat the case of homothety for \(m,l\neq 0\), let us go back form of \(\Psi\) \eqref{eqmic}. Taking the \(\rho\) and \(z\) derivatives of \eqref{eqmic} and setting to zero we obtain respectively

\begin{subequations}
\begin{align}
l\left(\gamma-\psi\right)_{,z\rho}+m\left(\gamma-\psi\right)_{,\rho\rho}&=0,\label{eqmicc1}\\
l\left(\gamma-\psi\right)_{,zz}+m\left(\gamma-\psi\right)_{,\rho z}&=0,\label{eqmicc2}
\end{align}
\end{subequations}
which can be combined to give the second order partial differential equation

\begin{eqnarray}\label{eqmicc3}
m^2\left(\gamma-\psi\right)_{,\rho\rho}-l^2\left(\gamma-\psi\right)_{,zz}=0,
\end{eqnarray}
which is clearly hyperbolic for \(l,m\neq0\). We shall look at specific examples of this case where the difference of the functions \(\gamma\) and \(\psi\) take on a particular form.

Consider first the case where \(\gamma-\psi\) separates as the linear sum

\begin{eqnarray}\label{eqmicc20}
\gamma-\psi=S+T,
\end{eqnarray}
where \(S=S\left(\rho\right)\) and \(T=T\left(z\right)\) (an obvious example is one where both \(\psi\) and \(\gamma\) split, i.e. where we may write \(\psi=h_1\left(\rho\right)+\xi_1\left(z\right)\) and \(\gamma=h_2\left(\rho\right)+\xi_2\left(z\right)\) for arbitrary functions \(h_1,h_2,\xi_1,\xi_2\)). Then \eqref{eqmicc3} becomes

\begin{eqnarray}\label{eqmicc21}
m^2S_{,\rho\rho}-l^2T_{,zz}=0,
\end{eqnarray}
from which we have the two equations

\begin{subequations}
\begin{align}
m^2S_{,\rho\rho}&=\lambda,\label{eqmicc22}\\
-l^2T_{,zz}&=\lambda,\label{eqmicc23}
\end{align}
\end{subequations}
for some constant \(\lambda\). The solutions to \eqref{eqmicc22} and \eqref{eqmicc23} are respectively

\begin{subequations}
\begin{align}
S&=\frac{\lambda}{m^2}\rho^2+c_1,\label{eqmicc24}\\
T&=-\frac{\lambda}{l^2}z^2+c_2,\label{eqmicc25}
\end{align}
\end{subequations}
for arbitrary constants \(c_1,c_2\neq 0\), and therefore 

\begin{eqnarray}\label{eqmicc26}
\gamma-\psi=\lambda\left(\frac{1}{m^2}\rho^2-\frac{1}{l^2}z^2\right)+c_1\rho + c_2z,
\end{eqnarray}
which gives 

\begin{eqnarray}\label{eqmicc27}
\Psi=2\lambda\left(\frac{1}{m}\rho-\frac{1}{l}z\right)+mc_1+lc_2,
\end{eqnarray}
where we have used \eqref{eqmic}. Hence we have that either \(\lambda=0\) or 

\begin{eqnarray}\label{hehe}
l\rho-mz=ml\bar{c},
\end{eqnarray}
for an arbitrary constant \(\bar{c}\), for the case of homothety. Using \eqref{eqq13} to substitute for \(\psi\) in \eqref{eqq14}, and comparing the result to \eqref{eqmicc26} we have the required forms of \(R\):

\begin{subequations}
\begin{align}
\lambda=0:\ R&=\frac{1}{k\alpha}e^{2\left[\bar{c}_1\rho+c_2z-r\left(z-\frac{l}{m}\rho\right)\right]},\label{eqmicc28}\\
l\rho-mz=ml\bar{c}:\ R&=\frac{1}{k\alpha\bar{c}_2}e^{\bar{m}\rho},\label{eqmicc29}
\end{align}
\end{subequations}
for constants

\begin{eqnarray*}
\begin{split}
\bar{c}_1&=\frac{1}{\pi m}+c_1;\\
\bar{c}_2&=e^{\left(mlc_2\bar{c}+\lambda\bar{c}^2+r\left(-mlc\right)\right)};\\
\bar{m}&=e^{\left(\frac{2}{\pi m}+c_1+\frac{l}{m}c_2+2\frac{\lambda}{m}\bar{c}\right)}.
\end{split}
\end{eqnarray*}
Hence, from \eqref{eqq20} \(\Psi\) becomes

\begin{subequations}
\begin{align}
\lambda=0:\ \Psi_{m,l\neq 0}&=\frac{1}{k\alpha}\left[m\bar{c}_1+lc_2+\bar{r}\right]-\frac{1}{\pi},\label{aamic1}\\
l\rho-mz=ml\bar{c}:\ \Psi_{m,l\neq 0}&=\frac{1}{2k\alpha}\bar{m}-\frac{1}{\pi},\label{aamic2}
\end{align}
\end{subequations}
where we defined 

\begin{eqnarray*}
\bar{r}=2\left(\frac{l}{m}-1\right)r\left(z-\frac{l}{m}\rho\right)=constant,
\end{eqnarray*}
i.e. \(r\left(z-\frac{l}{m}\rho\right)\) is constant. Notice that this means that we may write \eqref{eqmicc28} for the case \(\lambda=0\) as

\begin{eqnarray}\label{monster}
R_{\lambda=0}&=\frac{1}{k\alpha\bar{r}'}e^{2\left[\bar{c}_1\rho+c_2z\right]}\;,
\end{eqnarray}
where we have set

\begin{eqnarray*}
\bar{r}'=e^{\frac{\bar{r}m}{l-m}}.
\end{eqnarray*}
We also have the functions \(\psi\) and \(\gamma\) given  respectively by

\begin{subequations}
\begin{align}
\lambda=0:\ e^{2\psi}_{m,l\neq 0}&=\frac{1}{\bar{r}'}e^{2\left[\bar{c}_1\rho+c_2z\right]},\label{atmic1}\\
\lambda=0:\ e^{2\gamma}_{m,l\neq 0}&=\frac{1}{\bar{r}'\bar{r}^2}e^{2\left[\left(\bar{c}_1+c_1\right)\rho+c_2z\right]},\label{atmic2}\\
l\rho-mz=ml\bar{c}:\ e^{2\psi}_{m,l\neq 0}&=\frac{1}{\bar{c}}_2e^{\bar{m}\rho},\label{atmic3}\\
l\rho-mz=ml\bar{c}:\ e^{2\gamma}_{m,l\neq 0}&=\frac{1}{\bar{r}'\bar{c}_2^2}e^{2\left[\left(\bar{c}_1+c_1\right)\rho+c_2z\right]}.\label{atmic4}
\end{align}
\end{subequations}

Let us next consider the case where the difference \(\gamma-\psi\) factors as the product

\begin{eqnarray}\label{eqmicc30}
\gamma-\psi=ST,
\end{eqnarray}
where again \(S=S\left(\rho\right)\) and \(T=T\left(z\right)\) (an obvious example here is the case where both \(\psi\) and \(\gamma\) split as products, i.e. where we may write \(\psi=h_1\left(\rho\right)\xi_1\left(z\right)\) and \(\gamma=h_2\left(\rho\right)\xi_2\left(z\right)\), for arbitrary functions \(h_1,h_2,\xi_1,\xi_2\)). Then \eqref{eqmicc3} becomes

\begin{eqnarray}\label{eqmicc31}
m^2\frac{S_{,\rho\rho}}{S}-l^2\frac{T_{,zz}}{T}=0,
\end{eqnarray}
from which we have the two equations

\begin{subequations}
\begin{align}
S_{,\rho\rho}&=\frac{\lambda'}{m^2}S,\label{eqmicc32}\\
T_{,zz}&=\frac{\lambda'}{l^2}T,\label{eqmicc33}
\end{align}
\end{subequations}
for some constant \(\lambda'\). The solutions to \eqref{eqmicc32} and \eqref{eqmicc33} are respectively

\begin{subequations}
\begin{align}
\lambda'>0:\ \ S&=\bar{a}_1e^{\frac{\sqrt{\lambda'}}{m}\rho}+\bar{a}_2e^{-\frac{\sqrt{\lambda'}}{m}\rho},\label{eqmicc34}\\
T&=\bar{b}_1e^{\frac{\sqrt{\lambda'}}{l}z}+\bar{b}_2e^{-\frac{\sqrt{\lambda'}}{l}z},\label{eqmicc35}\\
\lambda'<0:\ \ S&=\bar{a}_1\cos\left(\frac{\sqrt{-\lambda'}}{m}\rho\right)+\bar{a}_2\sin\left(\frac{\sqrt{-\lambda'}}{m}\rho\right),\label{eqmicc36}\\
T&=\bar{b}_1\cos\left(\frac{\sqrt{-\lambda'}}{l}z\right)+\bar{b}_2\sin\left(\frac{\sqrt{-\lambda'}}{l}z\right),\label{eqmicc37}
\end{align}
\end{subequations}
for arbitrary constants \(\bar{a}_1,\bar{a}_2,\bar{b}_1,\bar{b}_2\), and therefore

\begin{subequations}
\begin{align}
\lambda'>0:\ \ \gamma-\psi&=\left(\bar{a}_1e^{\frac{\sqrt{\lambda'}}{m}\rho}+\bar{a}_2e^{-\frac{\sqrt{\lambda'}}{m}\rho}\right)\times\left(\bar{b}_1e^{\frac{\sqrt{\lambda'}}{l}z}+\bar{b}_2e^{-\frac{\sqrt{\lambda'}}{l}z}\right),\label{eqmicc38}\\
\lambda'<0:\ \ \gamma-\psi&=\biggl(\bar{a}_1\cos\left(\frac{\sqrt{-\lambda'}}{m}\rho\right)+\bar{a}_2\sin\left(\frac{\sqrt{-\lambda'}}{m}\rho\right)\biggr)\biggl(\bar{b}_1\cos\left(\frac{\sqrt{-\lambda'}}{l}z\right)\notag\\
&+\bar{b}_2\sin\left(\frac{\sqrt{-\lambda'}}{l}z\right)\biggr),\label{eqmicc39}
\end{align}
\end{subequations}
which gives the associated conformal factors as

\begin{subequations}
\begin{align}
\lambda'>0:\ \ \Psi&=2\sqrt{\lambda'}\left(a'e^{\frac{\sqrt{\lambda'}}{ml}\left(l\rho+mz\right)}+b'e^{-\frac{\sqrt{\lambda'}}{ml}\left(l\rho+mz\right)}\right),\label{eqmicc40}\\
\lambda'<0:\ \ \Psi&=\sqrt{-\lambda'}\biggl(a'\sin\left[\frac{\sqrt{-\lambda'}}{ml}\left(l\rho+mz\right)\right]+b'\cos\left[\frac{\sqrt{-\lambda'}}{ml}\left(l\rho+mz\right)\right]\biggr),\label{eqmicc41}
\end{align}
\end{subequations}
for arbitrary constants \(a'\) and \(b'\). Hence we must have that \(l\rho-mz=ml\bar{c}\) for either case \(\lambda'>0\) or \(\lambda'<0\) for the case of homothety. The difference \(\gamma-\psi\) in this case is therefore constant. Hence we have the required forms of \(R\) as

\begin{subequations}
\begin{align}
\lambda'>0:\ R&=\frac{1}{k\alpha}e^{\left[\left(\gamma-\psi\right)_{\lambda'>0}-\bar{r}\right]},\label{eqmicc42}\\
\lambda'<0:\ R&=\frac{1}{k\alpha}e^{\left[\left(\gamma-\psi\right)_{\lambda'<0}-\bar{r}\right]}.\label{eqmicc43}
\end{align}
\end{subequations}
The associated conformal factor in either case is simply, from \eqref{eqq20}

\begin{eqnarray}\label{eqmicc44}
\Psi=-\frac{1}{\pi}.
\end{eqnarray}

It is also not difficult to obtain the forms of the functions \(\gamma\) and \(\phi\), which are given by

\begin{subequations}
\begin{align}
\lambda'>0:\ e^{2\psi_{\lambda'>0}}&=e^{-2\bar{r}},\label{emica}\\
\lambda'>0:\ e^{2\gamma_{\lambda'>0}}&=e^{-4\bar{r}},\label{emicb}\\
\lambda'<0:\ e^{2\psi_{\lambda'<0}}&=e^{-2\bar{r}},\label{emicck}\\
\lambda'<0:\ e^{2\gamma_{\lambda'<0}}&=e^{-4\bar{r}},\label{emicd}
\end{align}
\end{subequations}
which implies they are constants. We therefore state the following propositions.

\begin{proposition}\label{lem203}
Let the metric \eqref{sav10} admit HKVs of the form \eqref{sav32}, where \(C_i=C_i\left(t,\rho\right)\) for \(i\in\lbrace{1,2,3,4\rbrace}\), and let \(C_2\) be constant. Then for \(C_2=m\neq0,C_3=l\neq 0\), the HKVs takes the form

\begin{eqnarray}\label{todee100}
\eta=\left(\pi_1t+\pi_2\right)\partial_t+m\partial_{\rho}+l\partial_{z}+\left(a_1-\alpha^2t\right)\partial_{\phi},
\end{eqnarray}
with \(\pi_1,\pi_2\) given in \eqref{hellothere}, where \(\pi,\alpha\neq 0\). 

If the difference \(\gamma-\psi\) splits as \eqref{eqmicc20}, then the associated conformal factor is given by \eqref{aamic1} or \eqref{aamic2}, and the functions \(R,\psi\) and \(\gamma\) are given by \eqref{monster} or \eqref{eqmicc29}, \eqref{atmic1} or \eqref{atmic3}, and \eqref{atmic2} or \eqref{atmic4}. 

On the other hand, if the difference \(\gamma-\psi\) factors as \eqref{eqmicc30}, then  the associated conformal factor is given by \eqref{eqmicc40} or \eqref{eqmicc41}, where the coordinates \(\rho\) and \(z\) satisfy \eqref{hehe}, and the functions \(R,\psi\) and \(\gamma\) are all constants given by \eqref{eqmicc42} or \eqref{eqmicc43}, \eqref{emica} or \eqref{emicck}, and \eqref{emicb} or \eqref{emicd}.

\end{proposition}

Now, let us consider the cases \(m=0,l\neq 0\) or \(m\neq 0,l=0\), in which case \eqref{eqmicc3} is parabolic. Then we have the metric function \(\gamma\) given respectively by

\begin{subequations}
\begin{align}
e^{2\gamma}_{m=0,l\neq 0}&=k^2\alpha^2R^2e^{-2\left(\frac{1}{\pi l}z-r_1\right)},\label{eqq21}\\
e^{2\gamma}_{m\neq 0,l=0}&=k^2\alpha^2R^2e^{-2\left(\frac{1}{\pi m}\rho-r_2\right)},\label{eqq22}
\end{align}
\end{subequations}
for arbitrary smooth functions \(r_1=r_1\left(\rho\right)\) and \(r_2=r_2\left(z\right)\). The functions \(C_i\) and \(\psi\) are given as in \eqref{eqq16} to \eqref{eqq19} and \eqref{eqq13}. The associated conformal factors are given respectively by

\begin{subequations}
\begin{align}
\Psi&_{m=0,l\neq 0}=-\frac{1}{\pi}+l\tilde{\alpha}\frac{R_{,z}}{R},\label{eqq23}\\
\Psi&_{m\neq 0,l=0}=-\frac{1}{\pi}+m\tilde{\alpha}\frac{R_{,\rho}}{R},\label{eqq24}
\end{align}
\end{subequations}
where we have defined

\begin{eqnarray*}
\tilde{\alpha}=k\alpha-\frac{1}{2k\alpha}.
\end{eqnarray*}

For the case with \(l=0\) and \(m\neq 0\), substituting \eqref{eqq13} into \eqref{eqq22}, and substituting the result into \eqref{eqmicc3}, and taking twice the \(\rho\) derivative, \eqref{eqmicc3} reduces to

\begin{eqnarray}\label{eqmicc4}
RR_{,\rho\rho}-R_{,\rho}^2=0,
\end{eqnarray}
whose solution is given by

\begin{eqnarray}\label{eqmicc5}
R=g_1e^{f_1\rho},
\end{eqnarray}
for arbitrary functions \(f_1=f_1\left(z\right)\) and \(g_1=g_1\left(z\right)\), which gives \eqref{eqq24} as

\begin{eqnarray}\label{eqmicc6}
\Psi_{m\neq0,l=0}=\frac{1}{\pi}+m\tilde{\alpha}f_1.
\end{eqnarray}
We therefore have that \(f_1\) must be constant for \(\Psi_{m\neq0,l=0}\) to be constant. 

Similarly, in the case that \(m=0\) and \(l\neq 0\) we have that \eqref{eqmicc3} becomes

\begin{eqnarray}\label{eqmicc7}
RR_{,zz}-R_{,z}^2=0,
\end{eqnarray}
whose solution is given by

\begin{eqnarray}\label{eqmicc8}
R=g_2e^{f_2\rho},
\end{eqnarray}
for arbitrary functions \(f_2=f_2\left(\rho\right)\) and \(g_2=g_2\left(\rho\right)\), which gives \eqref{eqq23} as

\begin{eqnarray}\label{eqmicc9}
\Psi_{m=0,l\neq 0}=\frac{1}{\pi}+l\tilde{\alpha}f_2.
\end{eqnarray}
Again, it is clear that \(f_2\) must be constant inorder for \(\Psi_{m=0,l\neq 0}\) to be constant. We therefore state the following

\begin{proposition}\label{lem201}
Let the metric \eqref{sav10} admit HKVs of the form \eqref{sav32}, where \(C_i=C_i\left(t,\rho\right)\) for \(i\in\lbrace{1,2,3,4\rbrace}\). If \(C_2\) is constant, then for \(C_3=l=0,C_2=m\neq 0\), the HKVs takes the form 
\begin{eqnarray}\label{todee101}
\eta=\left(\pi_1t+\pi_2\right)\partial_t+m\partial_{\rho}+\left(a_1-\alpha^2t\right)\partial_{\phi},
\end{eqnarray}
with \(\pi_1,\pi_2\) given in \eqref{hellothere}, where \(\pi,\alpha\neq 0\). The associated conformal factor is given by \eqref{eqmicc6}. The function \(\omega\) is  given by \eqref{eqq10} and \(\psi\) and \(\gamma\) are given respectively by

\begin{subequations}
\begin{align}
e^{2\psi}&=k\alpha g_1e^{f_1\rho},\label{live1}\\
e^{2\gamma}&=k^2\alpha^2g_1^2e^{-2\left[\left(\frac{1}{\pi m}-f_1\right)\rho-r_2\right]},\label{live2}
\end{align}
\end{subequations}
where \(g_1\) is a function of \(z\) and \(f_1\) is constant.
\end{proposition}

\begin{proposition}\label{lem202}
Let the metric \eqref{sav10} admit HKVs of the form \eqref{sav32}, where \(C_i=C_i\left(t,\rho\right)\) for \(i\in\lbrace{1,2,3,4\rbrace}\). If \(C_2\) is constant, then for \(C_2=m=0,C_3=l\neq 0\), the HKVs takes the form 

\begin{eqnarray}\label{todee102}
\eta=\left(\pi_1t+\pi_2\right)\partial_t+l\partial_{z}+\left(a_1-\alpha^2t\right)\partial_{\phi},
\end{eqnarray}
with \(\pi_1,\pi_2\) given in \eqref{hellothere}, where \(\pi,\alpha\neq 0\). The associated conformal factor is given by \eqref{eqmicc9}. The function \(\omega\) is  given by \eqref{eqq10}, and \(\psi\) and \(\gamma\) are given respectively by

\begin{subequations}
\begin{align}
e^{2\psi}&=k\alpha g_2e^{f_2\rho},\label{live1}\\
e^{2\gamma}&=k^2\alpha^2g_2^2e^{-2\left(\frac{1}{\pi m}z-f_2\rho-r_1\right)},\label{live2}
\end{align}
\end{subequations}
where \(g_2\) is a function of \(z\) and \(f_2\) is constant.
\end{proposition}

Finally, in the case that \(m\) and \(l\) are simultaneously zero, then \(M=\tilde{M}=C_2=C_3=\Psi=0\). Hence \eqref{sav32} are KVs of the form

\begin{eqnarray}\label{eqq25}
\eta=\pi_2\partial_t+a_1\partial_{\phi},
\end{eqnarray}

which allows us to state

\begin{proposition}\label{lem6}
Let the metric \eqref{sav10} admit HKVs of the form \eqref{sav32}, where \(C_i=C_i\left(t,\rho\right)\) for \(i\in\lbrace{1,2,3,4\rbrace}\). If \(C_2\) is constant, then for \(C_3=l=0,C_2=m=0\), the HKVs are KVs of the form \eqref{eqq25}.
\end{proposition}

We may also consider another scenario in which \(\eta\) is a KV: The case \(\gamma-\psi=c=constant\) (we will assume that \(m,l\neq 0\)). From \eqref{eqq9} we have

\begin{eqnarray}\label{eqq26}
l\psi_{,z}+m\psi_{,\rho}=-\beta,
\end{eqnarray}
which has the general solution

\begin{eqnarray}\label{eqq27}
\psi=-\frac{\beta}{m}\rho+r\left(l\rho-mz\right),
\end{eqnarray}
for an arbitrary smooth function \(r\left(l\rho-mz\right)\). We therefore have

\begin{subequations}
\begin{align}
e^{2\psi}&=e^{-2\left(\frac{\beta}{m}\rho-r\left(l\rho-mz\right)\right)},\label{eqq28}\\
e^{2\gamma}&=\tilde{c}e^{-2\left(\frac{\beta}{m}\rho-r\left(l\rho-mz\right)\right)},\label{eqq29}
\end{align}
\end{subequations}
where we have set \(\tilde{c}=e^{2c}\).

\subsection{The case \(W_{,\rho}-2W\left(\gamma-\psi\right)_{,\rho}=0\)}

The solution to \eqref{eq121} is given by

\begin{eqnarray}\label{eq125}
W=fe^{2\left(\gamma-\psi\right)},
\end{eqnarray}
where \(f=f\left(z\right)>0\) is an arbitrary function of \(z\). 

Now, substituting for \(\Psi\) and \(W\) into \eqref{eq109} using \eqref{eq108} and \eqref{micp} we obtain (after some simplification)

\begin{eqnarray}\label{michg1}
l\left(f_{,z}-2f\left(\gamma-\psi\right)_{,z}\right)=0.
\end{eqnarray}
Hence we either have that \(l=0\) or 

\begin{eqnarray}\label{michg2}
f_z-2f\left(\gamma-\psi\right)_{,z}=0.
\end{eqnarray}

We consider the two subcases in detail.

\subsubsection{\(f_{,z}-2f\left(\gamma-\psi\right)_{,z}=0\)}

First let us consider the case that \eqref{michg2} holds. Since \(f\) is a function of \(z\) only, then the solution to \eqref{michg2} is

\begin{eqnarray}\label{michg3}
f=\bar{k}e^{2\left(\gamma-\psi\right)},
\end{eqnarray}
for positive constant \(\bar{k}\), which implies \(\left(\gamma-\psi\right)_{,\rho}=0\), i.e. \(W\) is a function of \(z\) only, in which case

\begin{eqnarray}\label{redt1}
\Psi=l\left(\gamma-\psi\right)_{,z}.
\end{eqnarray}
We therefore have that \eqref{eq125} becomes

\begin{eqnarray}\label{michg4}
W=\bar{k}e^{4\left(\gamma-\psi\right)}.
\end{eqnarray}

Since the difference \(\left(\gamma-\psi\right)\) is independent of \(\rho\), one has the following form of \(\gamma\) and \(\psi\):

\begin{eqnarray}\label{todee}
\begin{split}
\gamma&=h_1\left(\rho\right)+\xi_1\left(z\right),\\
\psi&=h_2\left(\rho\right)+\xi_2\left(z\right);
\end{split}
\end{eqnarray}
for arbitraty functions \(h_1,h_2,\xi_1,\xi_2\), where \(h_1\) and \(h_2\) differ by a constant (possibly zero).

Comparing \eqref{eq102} and \eqref{eq107} we have (also using \eqref{micp})

\begin{eqnarray}\label{todee2}
-R^2C_{4,\rho}e^{-2\psi}=C_{2,t}e^{2\left(\gamma-\psi\right)}.
\end{eqnarray}
Substituting \eqref{michg4} into \eqref{eq107} we have (after rearrangement)

\begin{eqnarray}\label{todee3}
-C_{4,\rho}e^{-2\psi}=\frac{1}{\bar{k}}\omega C_{1,\rho}e^{-4\left(\gamma-\psi\right)},
\end{eqnarray}
which upon comparing with \eqref{todee2} gives

\begin{eqnarray}\label{todee4}
C_{2,t}=-\frac{1}{\bar{k}}\omega R^2C_{1,\rho}e^{6\left(\gamma-\psi\right)}.
\end{eqnarray}
Since \(C_2\) is a function of \(t\) only, \(C_1\) can be at most linear in \(\rho\), i.e.,

\begin{eqnarray}\label{todee5}
C_1=g\rho+q_1,
\end{eqnarray}
for a constant \(q_1\) and some function \(g=g\left(t\right)\). Also, we must have that 

\begin{eqnarray}\label{todee6}
\omega R^2e^{6\left(\gamma-\psi\right)}=q_2,
\end{eqnarray}
for a constant \(q_2\), in which case we have that \(\omega=\omega\left(z\right)\) and \(R\left(z\right)\). From \eqref{rhoeq} \(R\) takes the form

\begin{eqnarray}\label{todee7}
R=q_3z+q_4,
\end{eqnarray}
for constants \(q_3,q_4\). The component \(C_2\), from \eqref{todee4} therefore takes the form

\begin{eqnarray}\label{todee8}
C_2=-\frac{q_2}{\bar{k}}\int gdt,
\end{eqnarray}
where the constant \(q_2\) is specified by the functions \(\gamma,\psi\) and \(\omega\) (we will obtain \(\omega\) now) from \eqref{todee6}, while the component \(C_4\) becomes, from \eqref{todee3}

\begin{eqnarray}\label{todee9}
C_4=-\frac{\omega}{W}ge^{2\xi_2}\int e^{2h_2}d\rho.
\end{eqnarray}

Also, the function \(\omega\) is linear in \(z\). To see this, we use \eqref{eq101} and \eqref{micp} to simplify \eqref{eq104} as

\begin{eqnarray}\label{todee10}
C_{4,t}=-\frac{1}{k^2R^2}\left(\omega C_2h_{2,\rho}+l\omega_{,z}+4l\omega\xi_{2,z}-l\omega\xi_{1,z}\right).
\end{eqnarray} 
Since \(C_4\) is a function of \(t\) and \(\rho\) only and \(\omega\) is a function of \(z\) only, \(\omega_{,z}=\sigma_1\) must be constant which gives 

\begin{eqnarray}\label{todee11}
\omega=\sigma_1z+\sigma_2, 
\end{eqnarray}
for constants \(\sigma_1,\sigma_2\). Furthermore, we also have that \(\omega\xi_{1,z}=\bar{\sigma}\) and \(\omega\xi_{2,z}=\sigma'\), for constants \(\bar{\sigma},\sigma'\), from which we obtain \(\xi_1\) and \(\xi_2\) respectively as

\begin{subequations}
\begin{align}
\xi_1&=\frac{\bar{\sigma}}{\sigma_1}\ln\left(\sigma_1z+\sigma_2\right)+\bar{q},\label{todee12}\\
\xi_2&=\frac{\sigma'}{\sigma_1}\ln\left(\sigma_1z+\sigma_2\right)+q',\label{todee13}
\end{align}
\end{subequations}
for constants \(\bar{q},q'\). Hence, using \eqref{todee}we have

\begin{eqnarray}\label{todee14}
\left(\gamma-\psi\right)_{,z}=\frac{\Delta\sigma}{\sigma_1z+\sigma_2},
\end{eqnarray}
which gives \eqref{redt1} as

\begin{eqnarray}\label{todee15}
\Psi=l\frac{\Delta\sigma}{\sigma_1z+\sigma_2},
\end{eqnarray}
where we have set \(\Delta\sigma=\bar{\sigma}-\sigma'\). We see that \(\Psi\) vanishes if and only if \(\Delta\sigma=0\). This is true if the difference \(\Delta\xi=\xi_1-\xi_2\) is constant, for \(\omega\neq 0\). However, notice also that the case \(\omega=0\) gives \(\Delta\sigma=0\), as well as \(\Delta\xi=\xi_1-\xi_2\) being constant. From \eqref{todee15} it is quite easy to see that homothety requires \(\sigma_1=0\). Hence, \(\omega\), \(\xi_1\) and \(\xi_2\) must be constants. We can therefore state to the following

\begin{proposition}\label{lem100}
Let the metric \eqref{sav10} admit HKVs of the form \eqref{sav32}, where \(C_i=C_i\left(t,\rho\right)\) for \(i\in\lbrace{1,2,3,4\rbrace}\). If \(g_{\phi\phi}=\left(1/k^2\right)fg_{\rho\rho}\), then for \(f\) of the form \eqref{michg3}, \(\psi\) and \(\gamma\) split as \eqref{todee} and \(\omega\) is constant.The HKVs are of the form

\begin{eqnarray}\label{todee24}
\eta=\left(g\rho+q_1\right)\partial_t-\frac{q_2}{\bar{k}}\tilde{g}\partial_{\rho}+l\partial_z-gG\left(\rho,z\right)e^{2\xi_2}\partial_{\phi},
\end{eqnarray}
where we have defined the functions

\begin{eqnarray*}
\begin{split}
\tilde{g}\left(t\right)'&=\int g\left(t\right)dt,\\
G\left(\rho,z\right)&=\frac{\omega}{W}\int e^{2h_2}d\rho,
\end{split}
\end{eqnarray*}
for some function \(h_2=h_2\left(\rho\right)\) and constant \(\xi_2\). The associated conformal factor takes the form \eqref{todee15}, with \(\sigma_1=0\). We also have that \(\xi_1\) is constant.

\end{proposition}

\subsubsection{\(l=0\)}

Next we consider the case \(l=0\). This case gives (from \eqref{eq106})

\begin{eqnarray}\label{todee16}
\Psi=C_2\left(\gamma-\psi\right)_{,\rho}.
\end{eqnarray}

Combining \eqref{eq102} and \eqref{eq107} we have

\begin{eqnarray}\label{todee17}
C_{2,t}=\frac{1}{k^2}\frac{\bar{W}}{W}C_{1,\rho}e^{-2\left(\gamma-2\psi\right)}.
\end{eqnarray}
Again, \(C_2=C_2\left(t\right)\) and therefore we write

\begin{eqnarray}\label{todee18}
C_1=g'\rho+q'_1,
\end{eqnarray}
for some function \(g'=g'\left(t\right)\) and constant \(q'_1\). Furthermore, we must have

\begin{eqnarray}\label{todee19}
\frac{\bar{W}}{W}e^{-2\left(\gamma-2\psi\right)}=m',
\end{eqnarray}
for constant \(m'\), which gives

\begin{eqnarray}\label{todee20}
C_2=\frac{m'}{k^2}\int g'dt.
\end{eqnarray}
We obtain \(C_4\) from \eqref{eq107} as

\begin{eqnarray}\label{todee21}
C_4=-\frac{g'}{k^2}\int \omega e^{-2\left(\gamma-2\psi\right)}d\rho.
\end{eqnarray}
We therefore state the following

\begin{proposition}\label{lem8}
Let the metric \eqref{sav10} admit HKVs of the form \eqref{sav32}, where \(C_i=C_i\left(t,\rho\right)\) for \(i\in\lbrace{1,2,3,4\rbrace}\). If \(g_{\phi\phi}=\left(1/k^2\right)fg_{\rho\rho}\), then for \(l=0\), \(\psi\) and \(\gamma\) split as \eqref{todee} and \(\omega\) is constant.The HKVs are of the form

\begin{eqnarray}\label{todee24}
\left(g'\rho+q'_1\right)\partial_t+\frac{m'}{k^2}\tilde{g}'\partial_{\rho}-\frac{g'}{k^2}F\left(\rho,z\right)\partial_{\phi},
\end{eqnarray}
where we have defined the functions

\begin{eqnarray*}
\begin{split}
\tilde{g}'\left(t\right)&=\int g'\left(t\right)dt,\\
F\left(\rho,z\right)&=\int \omega e^{-2\left(\gamma-2\psi\right)}d\rho,
\end{split}
\end{eqnarray*}
for some function \(h_2=h_2\left(\rho\right)\) and constant \(\xi_2\), where \(\tilde{g}'\) and \(\left(\gamma-\psi\right)_{,\rho}\) are nonzero constants (\(\tilde{g}'\) being constant implies \(g'\left(t\right)=0\)). The associated conformal factor is given by

\begin{eqnarray}\label{todee23}
\Psi=\frac{m'}{k^2}\tilde{g}'\left(\gamma-\psi\right)_{,\rho}.
\end{eqnarray}
\end{proposition}

Indeed, an obvious solution for \(\gamma\) and \(\psi\) satisfying constant \(\left(\gamma-\psi\right)_{,\rho}\)  is

\begin{subequations}
\begin{align}
\gamma=h_3+\xi_3,\label{eqmicc442}\\
\psi=h_4+\xi_4,\label{eqmicc443}
\end{align}
\end{subequations}
where \(h_3=h_3\left(\rho\right),h_4=h_4\left(\rho\right)\) are linear polynomials in \(\rho\) and \(\xi_3=\xi_3\left(z\right),\xi_4=\xi_4\left(z\right)\) are arbitrary functions of \(z\).

\subsection{A particular case: \(\omega=0\)}

We consider the particular case where the metric \eqref{sav10} has no twist term, i.e. \(\omega=0\). This subclass of \eqref{sav10} contains the well studied Zipoy-Voorhees spacetimes \cite{zi1,vo1}, and the functions \(\psi\) and \(\gamma\) satisfy the following system

\begin{subequations}
\begin{align}
0&=\psi_{,\rho\rho}+\frac{1}{\rho}\psi_{,\rho}+\psi_{,zz},\label{amk1}\\
0&=\rho\left(\psi_{,\rho}^2-\psi_{,z}^2\right)-\gamma_{,\rho},\label{amk2}\\
0&=2\rho\psi_{,\rho}\psi_{,z}-\gamma_{,z}.\label{amk3}
\end{align}
\end{subequations}
In this case, from \eqref{eq104} and \eqref{eq107} it is either \(C_4=\bar{m}=constant\) or \(W=0\). We rule out the latter as \(\omega=0\implies \bar{W}=W\), in which case \(W=0\) implies that \(\underline{g}\) is noninvertible.

Using \eqref{eq102} we have

\begin{eqnarray}\label{todee22}
C_{2,t}=\frac{1}{k^2}C_{1,\rho}e^{-2\left(\gamma-2\psi\right)},
\end{eqnarray}
and again, as before we may write 

\begin{eqnarray}\label{todee23}
C_1=\bar{g}\rho+\bar{q}_1,
\end{eqnarray}
for some function \(\bar{g}=\bar{g}\left(t\right)\) and constant \(\bar{q}_1\). Also, we have that

\begin{eqnarray}\label{todee24}
e^{-2\left(\gamma-2\psi\right)}=k',
\end{eqnarray}
where \(k'\) is some constant. This gives \(C_2\) as

\begin{eqnarray}\label{todee25}
C_2=\frac{k'}{k^2}\int \bar{g}dt.
\end{eqnarray}

It can be shown that either \(\bar{g}\) is constant or the \(\rho=0\). To see this, notice that taking the \(\rho\) and \(z\) derivatives of \eqref{todee24} respectively gives 

\begin{subequations}
\begin{align}
\left(\gamma-\psi\right)_{,\rho}&=\psi_{,\rho}\label{todee26}\\
\left(\gamma-\psi\right)_{,z}&=\psi_{,z}.\label{todee27}
\end{align}
\end{subequations}
From \eqref{eq108} this gives

\begin{eqnarray}\label{todee28}
\Psi=C_2\psi_{,\rho}+l\psi_{,z},
\end{eqnarray}
and upon substituting \eqref{todee28} into \eqref{eq101} we have \(C_{1,t}=0\), which gives \(\bar{g}=const.\) or \(\rho=0\) (we shall assume that \(\rho\neq 0\)). 

Substituting \eqref{todee27} into \eqref{amk3} we obtain

\begin{eqnarray}\label{todee281}
2\psi_{,z}\left(\rho\psi_{,\rho}-1\right)=0.
\end{eqnarray}
Hence, either \(\psi=\psi\left(\rho\right)\) or 

\begin{eqnarray}\label{todee29}
\psi=\ln\rho + \zeta,
\end{eqnarray}
for some function \(\zeta=\zeta\left(z\right)\). If \(\psi=\psi\left(\rho\right)\), then from \eqref{amk2} we have that either \(\psi\) is constant (\(\implies \gamma\) is constant) or \(\psi=2\ln\rho+d_1\) (\(\implies \gamma=4\ln\rho+d_2\)) for constants \(d_1,d_2\). In the case that \eqref{todee29} holds, we can use \eqref{amk1} and it is not difficult to see that \(\zeta\) is linear in \(z\) which allows us to specify \(\psi\) and \(\gamma\) completely. We therefore have the following three possible configurations of the functions \(\psi\) and \(\gamma\): \(\psi\) is constant (\(\implies \gamma\) is constant); or \(\psi=2\ln\rho+d_1\) (\(\implies \gamma=4\ln\rho+d_2\)) for constants \(d',\tilde{d}\); or \(\psi=\ln\rho + d_3z+d'\) (\(\implies\gamma=2 d_3z+\tilde{d}\)), for constants \(d_1,d_2,d_3,d',\tilde{d}\). Furthermore, the general form of the CKVs is given by 

\begin{eqnarray}\label{todee30}
\eta=\left(\bar{g}\rho+\bar{q}_1\right)\partial_t+\frac{k'\bar{g}}{k^2}t\partial_{\rho}+l\partial_z+\bar{m}\partial_{\phi},
\end{eqnarray}
for some function \(\bar{g}=\bar{g}\left(t\right)\) and \(k',\bar{m},\bar{q}_1\) are constants. The associated conformal factor is given as follows:

\begin{enumerate}

\item If \(\psi\) is constant (\(\implies \gamma\) is constant), then

\begin{eqnarray}\label{todee31}
\Psi=0,
\end{eqnarray} 
in which case \(\eta\) is a KV.

\item If \(\psi=2\ln\rho+d_1\) (\(\implies \gamma=4\ln\rho+d_2\)), then

\begin{eqnarray}\label{todee31}
\Psi=\frac{2k'\bar{g}}{k^2\rho}t,
\end{eqnarray} 

\item If \(\psi=\ln\rho + d_3z+d'\) (\(\implies\gamma=2 \left(\ln\rho+d_3z\right)+\tilde{d}\)), then

\begin{eqnarray}\label{todee31}
\Psi=\frac{k'\bar{g}}{k^2\rho}t+ld_3.
\end{eqnarray} 

\end{enumerate}

Since the first case associates to trivial HKVs, i.e., the HKVs are KVs, only the last two cases associate to non-trivial HKV. We can therefore state the following

\begin{proposition}\label{lem110}
Let the metric \eqref{sav10}, with \(\omega=0\), admit CKVs of the form \eqref{sav32}, where \(C_i=C_i\left(t,\rho\right)\) for \(i\in\lbrace{1,2,3,4\rbrace}\). Then \eqref{sav32} are HKVs if neither \(\psi\) nor \(\gamma\) is constant, and \(\rho\) and \(\bar{g}t/\rho\) are constant with \(\rho\neq 0\) (if \(\rho=0\) then \(\bar{g}=0\) or \(t=0\) where we can rule out \(t=0\) as can be seen in \eqref{todee333} below). The form of the HKVs is given by 

\begin{eqnarray}\label{todee333}
\eta=\left(\bar{g}\rho+\bar{q}_1\right)\partial_t+\frac{k'\bar{g}}{k^2}t\partial_{\rho}+l\partial_z+\bar{m}\partial_{\phi}.
\end{eqnarray}
\end{proposition}

Notice that the case \(k=0\) can be avoided, since from \eqref{sav10} the resulting metric is just conformal to the real line, i.e., 

\begin{eqnarray*}
\underbar{g}=-e^{2\psi}dt^2.
\end{eqnarray*}

\section{Discussion}\label{soc6}

In this work, we have presented a detailed analysis for homothetic Killing vectors in the most general metric for SAV spacetimes. Our work has been restricted to analyzing the cases where the components of the homothetic Killing vectors are functions of the \(t\) and \(\rho\) coordinates only. As this is a very general consideration, this restriction simplified the conformal Killing equations and make them tractable. Once the existence of said homothetic Killing vectors is guaranteed, we have obtained the general forms that these vectors can take, along with their associated conformal factors. 

Firstly, it is seen that the component of any such homothetic Killing vector along the \(z\) direction - which is denoted by \(C_3\) - is necessarily constant. It is then further shown that a homothetic Killing vector must have either the component along the \(\rho\) direction - which is denoted by \(C_2\) -  being constant or, the metric components \(\underline{g}_{\rho\rho}\) and \(\underline{g}_{\phi\phi}\) are proportional in a specific way via \eqref{eq125}. In the case that \(C_2\) is constant, the forms of the functions \(\psi\), \(\gamma\) and \(\Psi\) (as well as whether \(\Psi\) is vanishing or non-constant function of the \(\rho\) and \(z\) coordinates) are dependent on the choices of the constants \(C_2\) and \(C_3\). On the other hand, if the metric functions \(\underline{g}_{\rho\rho}\) and \(\underline{g}_{\phi\phi}\) are proportional, then either the component \(C_3\) is identically zero, or there is further restriction on the proportionality factor between \(\underline{g}_{\rho\rho}\) and \(\underline{g}_{\phi\phi}\). The latter presents a case where the functions \(\psi\) and \(\gamma\) split as a sum of functions of \(\rho\) and \(z\). 

Obtaining the homothetic Killing vectors is reduced to solving a second order partial differential equation in \(\gamma-\psi\), which, given the possible configurations of the constants \(C_2\) and \(C_3\), are either parabolic or hyperbolic. The two parabolic cases where one of the constants vanishes was the easier to solve generally, and the required form of the function \(R\) was obtained. The results are presented in Proposition \ref{lem201} and Proposition \ref{lem202}.

For the hyperbolic case, we make some assumptions on the difference \(\gamma-\psi\): 

\begin{enumerate}

\item the first assumption is that the difference \(\gamma-\psi\) can be written as the linear sum of a function \(S\) of \(\rho\) only and a function \(T\) of \(z\) only; and

\item the second assumption is that the difference \(\gamma-\psi\) factors as a product of a function \(S\) of \(\rho\) only and a function \(T\) of \(z\) only.
\end{enumerate}
The results in the two cases are collected in Proposition \ref{lem203}.

We also considered the case where the constants \(m\) and \(l\) simultaneously vanish, in which case the homothetic Kiling vectors are trivial, the result which is collected in Proposition \ref{lem6}.

On the other hand, for the case where \(\underline{g}_{\rho\rho}\) and \(\underline{g}_{\phi\phi}\) are proportional, if the component \(C_3\) vanishes, then \(\tilde{g}'\) as defined in Proposition \ref{lem8}, and \(\left(\gamma-\psi\right)_{\rho}\) are both constants. If \(C_3=l\neq 0\) then there is further restriction on the proportionality factor between \(\underline{g}_{\rho\rho}\) and \(\underline{g}_{\phi\phi}\), in which case \(\psi\) and \(\gamma\) take the form \eqref{todee}. We also have that \(\omega\) is constant, and \(\xi_1\) and \(\xi_2\) are constants. 

In the case of vanishing \(\omega\), for non-constant \(\psi\) (and consequently \(\gamma\)), the result is collected in Proposition \ref{lem110}. Indeed, if \(\bar{g}\) is constant, then the coordinates \(t\) and \(\rho\) transform linearly as \(t\mapsto \alpha\rho\), for some arbitrary constant \(\alpha\). This transformation implies that the subclass of the metric \eqref{sav10} with vanishing rotation includes \(3\)-dimensional subspaces of \eqref{sav10}.

This work appears to be the first work to provide such considerations to HKV in the general metric \eqref{sav10}, and hence provides an incentive to further generalization and understanding some of the physical implications of the results herein. Another general problem to potentially consider would be investigating proper conformal Killing vectors in these spacetimes.

\section*{Acknowledgments}
RG is supported by National Research Foundation (NRF) of South Africa. SDM acknowledges that this work is based on research supported by the South African Research Chair Initiative of the Department of Science and Technology and the National Research Foundation. PKSD acknowledges support from the First Rand Bank, South Africa, and AS acknowledges support from the First Rand Bank, through the Department of Mathematics and Applied Mathematics, University of Cape Town, South Africa.

\end{document}